\documentclass[aps,a4paper,floatfix,nofootinbib]{revtex4}

\usepackage[utf8]{inputenc}
\usepackage{graphicx,color}
\usepackage{amsmath,amssymb,amsfonts,amsthm,amscd,bm}
\usepackage{booktabs}
\usepackage{cancel}

\def\bar{\overline}
\def\hat{\widehat}
\def\*{\star}
\def\[{\left[}
\def\]{\right]}
\def\({\left(}      

\def\){\right)}

\def\frac#1#2{\dfrac{#1}{#2}}
\def\inv#1{\dfrac{1}{#1}}

\def\d{\partial}

\def\2pi{\hbox{$2\pi i$}}

\def\dsl{\raise.15ex\hbox{/}\kern-.57em\partial}
\def\Dsl{\,\raise.15ex\hbox{/}\mkern-.13.5mu D}
\def\vep{\varepsilon}

      \def\CC{{\cal C}}

      \def\CO{{\cal O}}

\def\2pi{\hbox{$2\pi i$}}

\def\dsl{\raise.15ex\hbox{/}\kern-.57em\partial}
\def\Dsl{\,\raise.15ex\hbox{/}\mkern-.13.5mu D}
\font\numbers=cmss12
\font\upright=cmu10 scaled\magstep1
\def\stroke{\vrule height8pt width0.4pt depth-0.1pt}
\def\topfleck{\vrule height8pt width0.5pt depth-5.9pt}
\def\botfleck{\vrule height2pt width0.5pt depth0.1pt}
\def\Zmath{\vcenter{\hbox{\numbers\rlap{\rlap{Z}\kern
    0.8pt\topfleck}\kern 2.2pt
    \rlap Z\kern 6pt\botfleck\kern 1pt}}}
\def\Qmath{
    \vcenter{\hbox{\upright\rlap{\rlap{Q}\kern3.8pt\stroke}\phantom{Q}}}}
\def\Nmath{\vcenter{\hbox{\upright\rlap{I}\kern 1.7pt N}}}
\def\Cmath{\vcenter{\hbox{\upright\rlap{\rlap{C}\kern
                   3.8pt\stroke}\phantom{C}}}}
\def\Rmath{\vcenter{\hbox{\upright\rlap{I}\kern 1.7pt R}}}
\def\Z{\ifmmode\Zmath\else$\Zmath$\fi}
\def\Q{\ifmmode\Qmath\else$\Qmath$\fi}
\def\N{\ifmmode\Nmath\else$\Nmath$\fi}
\def\C{\ifmmode\Cmath\else$\Cmath$\fi}
\def\R{\ifmmode\Rmath\else$\Rmath$\fi}
\def\barray{\begin{eqnarray}}
\def\earray{\end{eqnarray}}
\def\beq{\begin{equation}}
\def\eeq{\end{equation}}

\def\n{\noindent}

\def\smallhalf{{\scriptstyle \inv{2}}}

\def\smallhalf{\tfrac{1}{2}}

\def\AA{\leavevmode\setbox0=\hbox{h}
\dimen0=\ht0 \advance\dimen0 by-1ex\rlap{\raise.67\dimen0\hbox{\char'27}}A}

\def\Arg{{\rm Arg}\,}

\makeatletter
\def\iddots{\mathinner{\mkern1mu\raise\p@
\vbox{\kern7\p@\hbox{.}}\mkern2mu
\raise4\p@\hbox{.}\mkern2mu\raise7\p@\hbox{.}\mkern1mu}}
\makeatother

\theoremstyle{plain}

\theoremstyle{remark}

\def\arg{{\rm arg}}
\def\Arg{{\rm Arg}}

\def\n{\noindent}

\def\thetasum{\theta_S}

\def\ELandau{E_L} 

\def\Ec{E^c}
\def\Bc{B^c}

\def\rhozero{\rho_\bullet}
\def\tzero{t_\bullet}
\def\sigmazero{\sigma_\bullet}

\def\ncal{\mathfrak{n}}

\def\Temp{{\bf T}}

\begin{document}

\title{Phenomenological formula for Quantum Hall resistivity \\
 based on the Riemann zeta function}
\author{
  Andr\'e  LeClair\footnote{andre.leclair@gmail.com} 
}
\affiliation{Cornell University, Physics Department, Ithaca, NY 14850, USA}

\medskip

\begin{abstract}
We propose a formula constructed out of elementary functions that captures many of the detailed features of the transverse
resistivity $\rho_{xy}$ for the integer quantum Hall effect.   It is merely a phenomenological formula in the sense that it is not based on any 
transport calculation for  a specific   class of physical  models involving electrons in a disordered landscape,   thus,  whether a physical model exists which realizes this resistivity remains an open question.   Nevertheless,  since the formula involves the Riemann zeta function and its non-trivial zeros play a central role,   it is amusing to consider the implications of the Riemann Hypothesis in light of it.   

\end{abstract}

\maketitle

\vskip -0.8 truecm

\tableofcontents

\section{Introduction}

\begin{figure}[t]
\centering\includegraphics[width=1.0\textwidth]{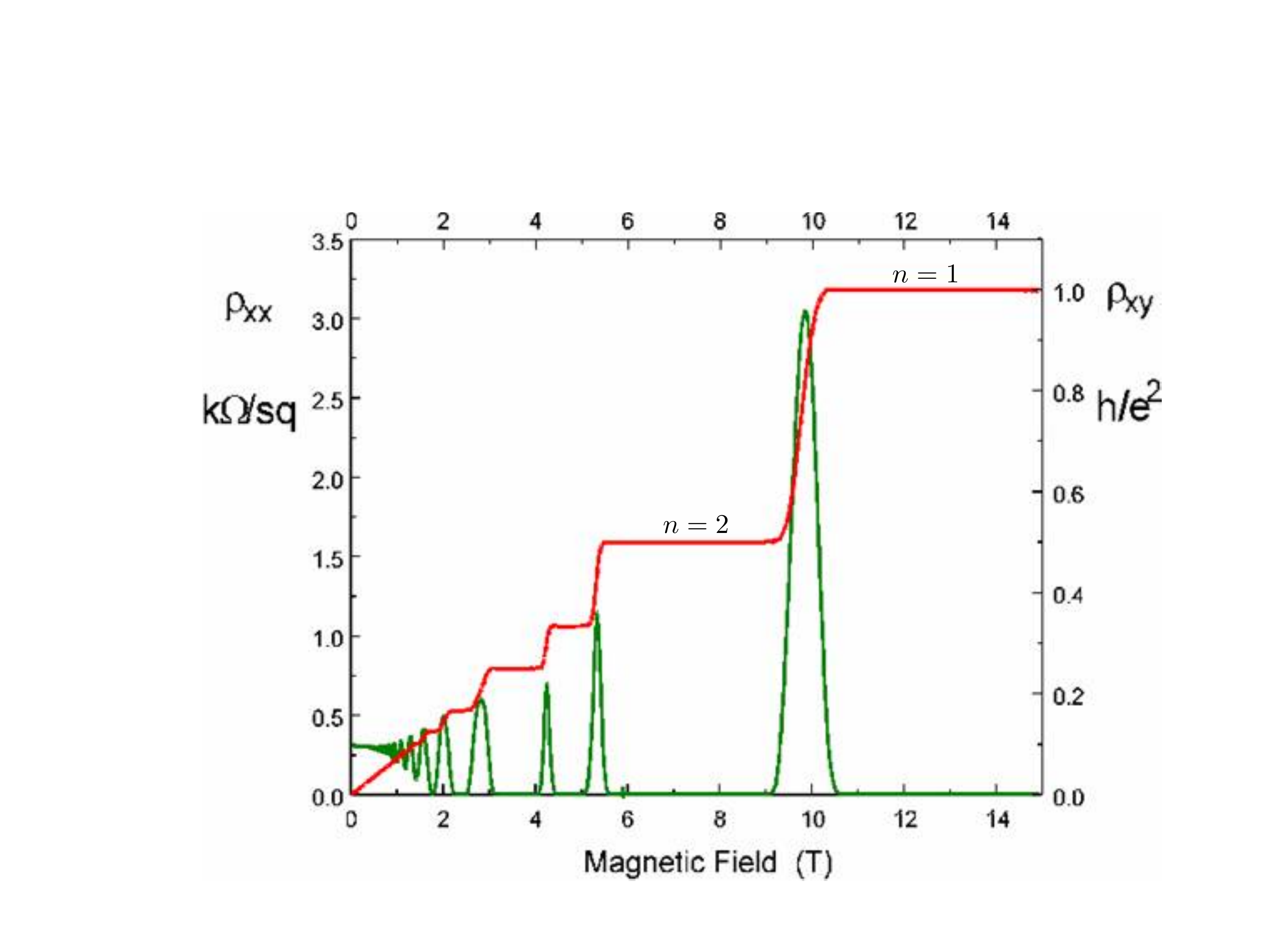}
\caption{Experimental data on the Hall resistivity $\rho_{xy}$ as a function of magnetic field at very low temperatures (red curve, 
the green curve is the  longitudinal $\rho_{xx}$.) }
 \label{Data}
\end{figure}

The integer quantum Hall effect (IQHE) refers to the remarkable properties of a non-interacting 
electron gas in two spatial dimensions in the presence of a magnetic field $B$ and a disordered potential.  
   These properties are only observed under the extreme conditions of 
strong magnetic field and very low temperature which explains why it was discovered relatively late \cite{vonKlitzing}.   
Figure \ref{Data} shows some typical experimental data.\footnote{We found this image on the internet but could not locate 
the article which  presented this figure, which are numerous.} 
The most striking feature is the exact quantization of the transverse resistivity on the ``plateaux": 
\beq
\label{rhoxy}
\rho_{xy} = \inv{n} \, \frac{h}{e^2}, ~~~~n = 1, 2, 3, \ldots
\eeq
where $h$ is Planck's constant and $e$ the electron charge.\footnote{The exact quantization $n= 1,2,3, \ldots$  has been verified experimentally to
within $10^{-10}$.}

The theoretical understanding of the IQHE is by now well-developed and there are many excellent reviews and books \cite{Girvin}. 
Some early pioneering works were by Laughlin and Halperin \cite{Laughlin,Halperin}.    Let us review some of the very basics. 
The pure problem of a single electron in a magnetic field was solved long ago by Landau.    
The eigenstates fall into Laudau levels with energy $\ELandau (n) = ( n + \smallhalf ) \hbar e B/m$ where $m$ is the electron mass.  
Each Landau level has a large degeneracy of states $N_{{\rm states}} = \Phi/ \Phi_0$ where $\Phi$ is the magnetic flux and $\Phi_0 = h c/e$.  
 These are all de-localized states. 
The IQHE is instead a many body problem,  namely a finite density electron gas at finite temperature  with non-zero $B$, 
{\it in the presence of disorder,}  such as various kinds of random  impurities or defects.       The plateaux would not exist without this disorder 
and a complete understanding involves Anderson localization.     Since we will not be considering any detailed physically motivated 
hamiltonian,  
let us just give a rough  and brief picture of the phenomenon.  At  energies  approximately in the gap between the pure  Landau levels the states are localized  due to disorder and thus don't contribute to the conductivity and it remains constant on the plateaux.  At certain critical  energies $\Ec_n$ near the original Landau levels, where by definition 
$\Ec_{n+1} > \Ec_n$,   there is a quantum phase transition where some states are delocalized and contribute to a change of  the conductivity.     
Let us write this in terms of the Fermi energy $E_F$:
\beq
\label{EFprop}
{\rm If} ~~ \Ec_{n} > E_F > \Ec_{n-1} ~~~~{\rm then } ~~ \rho_{xy} =  \inv{n} \, \frac{h}{e^2}
\eeq
For the experiments it  is easier to control the magnetic field,  thus we define critical $\Bc_n$ where $\Bc_{n+1} < \Bc_n$ and:
\beq
\label{Bprop}
{\rm If} ~~ \Bc_{n} < B <  \Bc_{n-1}  ~~~~{\rm then } ~~ \rho_{xy} =  \inv{n} \, \frac{h}{e^2} .
\eeq

It will also be important to point out that the integer $n$ in the quantization of $\rho_{xy}$ was eventually understood as a topological invariant, 
sometimes referred to as a Chern number \cite{Thouless}.    This fact explains the robustness of the quantized plateaux in spite of 
variable details of the sample,  in particular the realization of disorder,   which varies from sample to sample.   
It is also known that the transitions between plateaux are infinitely sharp at zero temperature,  and for the most part we assume we are in this situation.
We wish to also mention that the IQHE also exists in models without Landau levels \cite{Haldane}.

Although the theory of the IQHE is well-developed,   the explicit calculation of $\rho_{xy}$ is a difficult problem since it is a transport property in the presence of disorder.    One can attempt disorder averaging,  however this is also notoriously difficult,  especially at the critical transitions.     In fact,  the precise nature of the quantum critical points at the transitions remains unknown;   even the delocalization exponent is only known numerically.    We will consider this exponent in Section IV.   
Inspection of Figure \ref{Data} indicates many complicated details,  such as the following.    The widths of the plateaux 
vary considerably for a single sample,  becoming smaller as $B$ is decreased.    Apart from this,   the critical values $\Bc_n$ appear random,
and certainly depend on the realization of the disorder,   of which there are infinite possibilities.    The resistivity vanishes at $B=0$,  where it  is approximately linear.

In this paper,  we will merely present an explicit  mathematical function that has many of the same properties as the measured 
$\rho_{xy}$.   It is not such a simple matter  to conjure up such a function,  since the measured $\rho_{xy}$ has many detailed features.
   Our proposed formula is relatively simple, being constructed from the gamma and zeta functions, and captures many of  these  salient features.     Hence our terminology ``phenomenological",  since we don't attempt  to compute the resistivity for  any specific many-body hamiltonian,  which in any case is a very difficult problem due to the disorder.      For these reasons,   it will be left as an open question as to whether any specific model hamiltonian
for the IQHE has the transport properties that correspond to our formula,  even approximately.  
We wish to point out however that the zeta function has an integral representation involving the Fermi-Dirac distribution
$1/(e^\vep +1 )$ which we recall in the Appendix \eqref{IntRepFermi},   and this provides some encouragement 
that our formula may eventually be understood as a resistivity for a gas of fermions.   
Clearly some guesswork is involved and this exercise will not necessarily prove to be fruitful since it is not based on methods used to compute transport,  such as a Kubo formula.      
However we wish to point out that this kind of guesswork has been proven to be successful for some well-known  problems. 
For instance,  let us just mention Veneziano's guess of a scattering amplitude expressed only in terms of ratios of gamma functions led to the development of the huge field of string
theory\cite{Veneziano}.   Of course we cannot promise such success here,  however we feel it is still worth pursuing.   

For our proposed formula,  the non-trivial zeros of the zeta function play a crucial role,  and thus can perhaps provide some new insights on
the Riemann Hypothesis (RH).   As emphasized above,  we do not know if any microscopic quantum many body model leads to a resistivity 
that corresponds to our phenomenological formula.    Nevertheless,    if we hypothesize  such a connection,   then the RH can be interpreted in light of it.      There are very few approaches to the RH based essentially  on physics,   and this may be new one. 
We should mention approaches based on the Hilbert-P\`olya idea that perhaps there exists a single particle  quantum hamiltonian whose eigenvalues 
are equal to the ordinates of zeros on the critical line.     This has been pursued by Berry,  Keating,  Sierra and others \cite{BerryK,BerryK2,Sierra}.
A very different approach is based on ideas in statistical mechanics,  in particular properties of random walks, where the randomness arises from the pseudo-randomness of the prime numbers.   
See in particular the most recent work  \cite{Mussardo} and references therein.      
To our knowledge a possible connection between the IQHE and the RH has not been explored before,  and our hope is  that this short
work may shed some light on at least one of them.    We wish to mention though that a relation between the pure Landau problem and the Riemann zeros was proposed in \cite{SierraTownsend},  however without disorder this is not the same physics as the IQHE which involves transport rather than energy eigenvalues.

\section{A phenomenological formula for $\rho_{xy}$}

Let us straightaway present our formula.   Let $s=\sigma + i t$ be a complex variable and first define  the real function:
\beq
\label{thetaDef}
\theta(\sigma, t) = \Im \log \Gamma (s/2)  - t \, \log \sqrt{\pi }+ \arg \, \zeta(s), ~~~~~(s = \sigma + i t).
\eeq
First a few words about the components of this function,   which can in fact be interpreted as an angle, as we will explain.        
$ \Im \log \Gamma\( \smallhalf (\sigma+it) \)$ should be understood as  $\arg \, \Gamma (s/2)$.\footnote{In Mathematica LogGamma[z] is a built in function.}    
It is important here and elsewhere that in the above equation it is  $\arg$ and not $\Arg$,  where the latter  by definition is the principal branch.    
Remind that for any meromorphic function $f(s)$,  away from a zero or pole one  can always calculate
$\Arg \, f(s) = \arctan \, \Im f(s)/\Re f(s)$ on the principal branch $-\pi < \Arg < \pi$.   
On the other hand,  $\arg$ keeps track of branches and needs to be defined differently,   typically using piecewise integration 
of $f'(s)/f(s)$ from some known point.  ($ \arg (e^{i \alpha}) = \alpha$ for $-\infty < \alpha < \infty$.)  For instance for $\arg \, \zeta(\smallhalf
 +i t)$ one standard  definition  is via the contour $\CC'$ in Figure
\ref{argFigure}.    
In summary, if $s$ is neither a zero nor a pole,  then  
\beq
\label{args}
\arg \, f(s) = \Arg \, f(s) ~~~{\rm mod} ~ 2 \pi,   ~~~~~~ -\pi < \Arg\, f(s) < \pi,
\eeq
however in the present context,  the ${\rm mod} ~ 2\pi$ will be important.  
The contribution from  $\arg \, \Gamma$ is a smooth function that grows with $t$.    Using the Stirling formula and its corrections one can easily show for large $t$:
\beq
\label{RS}
\Im \log \Gamma \( \smallhalf (\sigma+it)\) - t \, \log \sqrt{\pi } = 
\frac{t}{2} \log \( \frac{t}{2\pi e}\) + \frac{\pi (\sigma -1)}{4}  - \frac{(3 \sigma^2 - 6 \sigma +2)}{12 t} + \CO(1/t^3) .
\eeq
On the other hand,  $\arg \, \zeta (s)$ is a much more complicated quantity and where all the action is. 
  Some basic facts we need about the zeta function 
are collected in the Appendix.      There is no analog of Stirling's formula that would lead to anything as simple as \eqref{RS}.     An important role will be played by $S(t) = \arg \, \zeta (\smallhalf+it) / \pi$,  and there is a large mathematical literature concerning it.   Here let us just mention Selberg's central limit theorem \cite{Selberg}:   $S(t)$ satisfies a normal distribution with zero mean 
and standard deviation  equal to$\tfrac{1}{\sqrt{2} \pi} \sqrt{ \log \log t}  $ in the limit $t\to \infty$.    Thus it grows very,  very slowly with $t$ compared to the first terms in \eqref{thetaDef}.  Thus to a high probability,  $\arg \, (\smallhalf + it)$ is nearly always on the
 principal branch.\footnote{For instance,  from Selberg's theorem one deduces that in the range $0<t<1000$ the probability that
 $\arg \, \zeta(\smallhalf + it )$ is not on the principle branch is only $0.0014$.  In practice this implies that in many cases one can use 
 $\Arg$ in Mathematica to compute $\arg$,  however this should be done with some care.}
    Nevertheless it can attain infinitely large values,  though very  rarely,  due to the tail of the normal distribution.

\begin{figure}[t]
\centering\includegraphics[width=.7\textwidth]{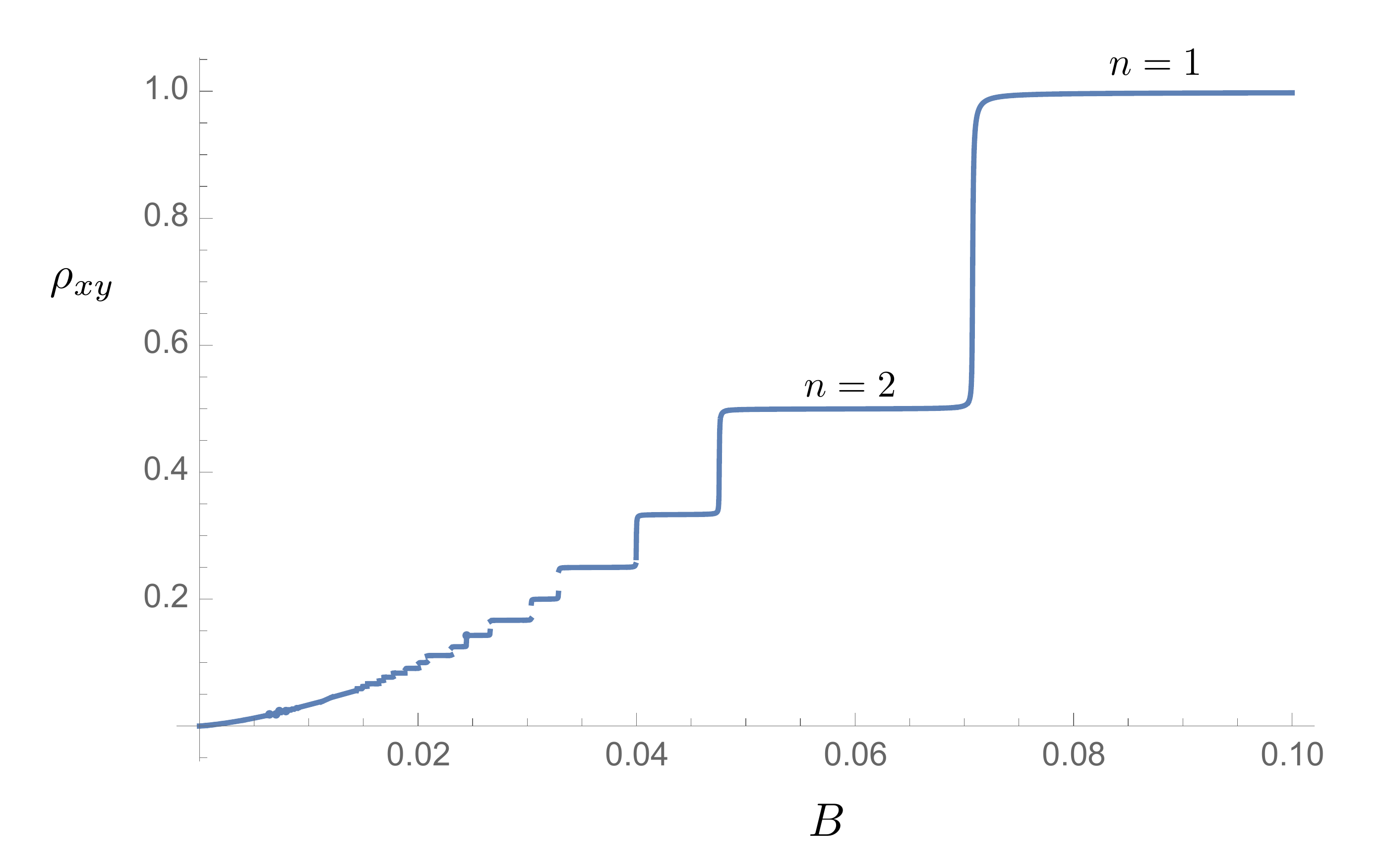}
\caption{Plot of the function \eqref{rhoxy}.  We have shifted $\sigma = \smallhalf  \to \smallhalf + \delta$ with 
very small positive $\delta$ in order to smooth out the transitions.   At $\sigma = \smallhalf$,  the transitions are infinitely sharp 
as expected at zero temperature.}
 \label{rhoxyPlot}
\end{figure}

With these preliminaries,   we present  the phenomenological function we advertised.   
Henceforth we work in units where $h=c=1$,  we set $m=e=1$,  and $B$ is expressed in dimensionless units,   such as $B/B_0$ where
$B_0$ is a reference magnetic field strength.   
We first present the simplest version,  
and defer presenting some deformations to Section IV: 
\beq
\label{rhoxyformula}
\rho_{xy}  (B) = \frac{\pi}{\theta(\smallhalf, 1/B) + 2 \pi } .
\eeq
In Figure \ref{rhoxyPlot} we plot the above function,  and it certainly shows  some resemblance to the data in Figure \ref{Data}.

For fixed $B$,  the Fermi energy is a function of the number density of electrons.   If the density is fixed,  as in a specific sample,  then the Fermi energy depends on $B$,   thus one can observe the plateaux by either varying $E_F$ or $B$.  The relation between these quantities 
can be complicated.    Note however that since the degeneracy of each  pure Landau level is proportional to $B$,  we roughly expect that
$E_F \propto 1/B$.\footnote{In two spatial dimensions one expects $E_F \propto \rho$ where $\rho$ is the number density of electrons. For instance,   in natural units with $\hbar = c= 1$, a  dimensionally correct relation is $E_F \propto \(\frac{\rho}{me \ell^2 } \) \inv{B}$  where $\ell$
a length scale   characterizing the disorder.   The present work does not rely on the specific relation between $E_F$ and $B$.} 
Since we will not deal with any specific material,   and we are not experts,  as an oversimplification we will simply {\it define}  for now 
\beq
\label{EBprop} 
\Ec_n  \equiv  1/\Bc_n
\eeq
in appropriate units.    Variations of the above formula will be considered in Section IV.  
  
  \bigskip
Let us list some of the main features of \eqref{rhoxyformula}:

\bigskip
$\bullet$ ~~ The values of the resistivity are exactly quantized as $\rho_{xy} = 1/n$  on the plateaux. 
{\it If one assumes the RH, and all the non-trivial zeros are simple}, 
 then this  is explained by the well-known result that the number of zeros  of $\zeta (s)$ inside the critical strip 
 $0<\sigma<1$ with $0<t<T$ is given by
\beq
\label{NofT}
N(T) = \theta(\smallhalf, T)/\pi + 1 .
\eeq
This is a consequence of  Cauchy's argument principle, and is reviewed in the Appendix.

\bigskip
$\bullet$ ~~ Let $\rho_n = \smallhalf + i t_n$ denote the $n$-th zero of $\zeta$ on the upper critical line,  where by convention
$n=1,2, \ldots$ and the first few are $t_1 = 14.134.., ~ t_2 = 21.022....$.   
Again assuming the RH,  the critical values are given precisely by these zeros:
\beq
\label{BcEc}
\Bc_n = 1/t_n ~~ \Longrightarrow ~~ \Ec_n = t_n  .
\eeq

\bigskip
$\bullet$  ~~ The widths of the plateaux,  $\Bc_n - \Bc_{n+1}$,  become smaller and smaller as $B$ is decreased,  
which resembles the experimental data.     
Although we did not attempt a fit to the data in Figure \ref{Data},  one can check that the ratios $\Bc_{n+1}/\Bc_n$ are roughly  equal to 
$t_n/t_{n+1}$.  

\bigskip
$\bullet$ ~~ The resistivity vanishes at $B=0$  nearly linearly:
\beq
\label{Bzero}
\lim_{B \to 0} \rho_{xy} (B) \approx \frac{2 \pi B}{ \log(1/ B) } .
\eeq
By replacing $B$ with $B |\log (B)| $ in \eqref{rhoxyformula},   $\rho_{xy} (B)$ is linear near $B=0$ up to much smaller  $\log \log B$ corrections. We will consider
other kinds of deformations in Section IV.

\section{Interpretation in connection with the Riemann Hypothesis}

The RH is the conjecture that all zeros inside the critical strip $0<\sigma <1$ are on the critical line $\sigma = \smallhalf$. 
Whether these zeros are simple or not also remains a difficult  open problem.     
In the last section we already commented that  our phenomenological formula \eqref{rhoxyformula} can only describe the IQHE if 
the RH is true and all zeros are simple.      Let us elaborate.  
Suppose the RH is false such that there are zeros  $\rhozero$ off the line,   which necessarily come in pairs 
$\rhozero = \sigmazero + i \tzero,  1- \sigmazero + i \tzero$.    These zeros contribute to $N(T)$ with multiplicity $2$ or more.   
This would imply that the transverse conductivity $1/\rho_{xy}$  does not always jump by $1$ at the transitions,   but rather sometimes jumps
 by $2$, or more depending on their multiplicity.  
In other words not all integers $n$ in \eqref{rhoxy} would be  physically realized.      Furthermore,  
assuming the RH,  if the zeros were not simple,   then $n$ would not always jump by exactly $1$,  but rather by the order of the zero,  and again some $n$ would be absent.  

In our  simplified physical picture thus far,  the critical energies $\Ec_n$ are identified as the Riemann zeros $t_n$ based on \eqref{EBprop}.    
Whereas the quantization $n$ in \eqref{rhoxy} is robust,  i.e. sample independent due to its relation to the Chern topological number,
the values at the transition $\Ec_n$ are not universal.  They depend in particular on the realization of the disorder,  i.e. details of the 
disorder, among other properties of the sample.      
A famous conjecture is that the differences between Riemann zeros $t_n$ satisfies the GUE statistics of random hamiltonians \cite{Montgomery,Odlyzko}.     In the present context this randomness is naturally explained as arising from the random disorder.  
In fact random matrix theory has  already been applied to quantum transport in disordered systems \cite{Beenakker}.
The actual Riemann zeros must then correspond to a special realization of disorder,  just as the Hilbert-P\'olya idea involves some as yet unknown hamiltonian. 
It is important then that our formula \eqref{rhoxyformula} can be deformed in order to accommodate for variable 
$\Ec_n \neq t_n$,  and other variations such as finite temperature.       This will be explored in Section IV.    
  It is interesting to note that extreme values of $|\zeta ( \smallhalf  + i t)|$  were studied from the perspective of the so-called freezing transition in disordered landscapes \cite{Fyodorov1,Fyodorov2},   and the latter is an important component of the physics of the  IQHE.

Returning to pure mathematics,  the angle $\theta(\sigma, t)$ contains all the information about the zeros although  in a not so transparent way.   
This was developed in a precise manner in \cite{Gui}.   
Referring to  the completed zeta function $\chi (s)$ in \eqref{chidef},  it is known that inside the critical strip $\chi (s)$ and $\zeta (s)$ have the same zeros.   
Obviously at a zero $\rho_n$,  the modulus $| \chi (\rho_n)| =0$,  however such a formula is not very useful in enumerating the zeros.   
Assuming the RH,  one can in fact extract the actual exact zeros $t_n$  from $\theta (\sigma, t)$ in the following way.    
First it is important that even though $| \chi (s) | =0$ at a zero,  its argument $\theta(\sigma, t)$ is still well defined {\it once one specifies a contour indicating the direction of approach to the zero}.     Let us provide a slightly different  argument than  the one presented in \cite{Gui}.\footnote{This version was found during discussions with Giuseppe Mussardo.}   
On the critical line $\sigma = \smallhalf$,  by the functional equation \eqref{funceqn},  $\chi(s)$ is real.    If the RH is true and the zeros are simple, 
 it must simply change sign at each zero.   Thus, assuming the RH and the simplicity of the zeros,  $\theta(\smallhalf, t)$ must jump by $\pi$ at each zero.    Approaching zeros along the critical line from below, one
finds 
\beq
\label{frombelow}
\lim_{\epsilon \to 0^+} \, \theta(\smallhalf, t_n - \epsilon ) = \pi (n-1).
\eeq
Rotating counterclockwise by $\pi/2$,  one deduces 
\beq
\label{fromright}
\lim_{\delta  \to 0^+} \, \theta(\smallhalf + \delta,   t_n  ) = \pi (n- \tfrac{3}{2} ) .
\eeq
The above equation was used in \cite{Gui} to calculate zeros to very high accuracy.\footnote{To our knowledge the above equation was first proposed by the author.  See \cite{Gui} and references therein.}      
It fact it was proven that if there is a unique solution to \eqref{fromright} for every $n$,  then the RH is true and 
all zeros are simple.  
The importance of approaching the zeros from the right of the critical line is two-fold.  
For so-called $L$-functions based on non-principal Dirichlet characters,  there are strong arguments that 
the Euler product formula (EPF) converges for $\sigma>\smallhalf$ (See \cite{LM1}).   For zeta itself,  the Euler product formula is 
\beq
\label{EPF}
\zeta (s) = \prod_p \( 1 - \inv{p^s} \)^{-1} 
\eeq
where $p$ is a prime, and  the EPF converges only for $\sigma > 1$,  unlike what is expected for $L$-functions based on 
non-principal Dirichlet characters.   
 To the left of the critical line,   $\arg \, \zeta (s)$ behaves 
quite differently than from the right:   from  equation \eqref{argrelations} one sees that to the left there are many more changes of branch compared with  to the right.       In conjunction with the EPF,  one can use the latter to approximate 
$\arg \, \zeta (\smallhalf + i t)$ and thereby compute the zeros $t_n$ directly from the prime numbers with a truncation of the EPF, at least approximately \cite{ALzeta}.

\def\zexp{z}

\section{Deformations of the  formula for $\rho_{xy}$:  variable $\Bc_n$ and modeling finite temperature}

As explained above,   we interpreted the critical $\Ec_n$ as equal to the exact Riemann zeros on the critical line.  
In reality,  for a specific experimental sample,   these $\Ec_n$  are not equal to the exact Riemann zeros  $t_n$.    It is thus important for our proposal that 
the $\Ec_n$ can be deformed away from the exact and known $t_n$ without spoiling the exact quantization of $n$ in \eqref{rhoxy}.   One expects this is possible since $n$ is topological.        This can be done in many ways,  and in this section we explore a few.  

\bigskip
\n $\bullet$ ~~ {\bf Deforming the functional dependence on $B$.} ~~
The most straightforward deformation is to change the function of $B$ inside $\theta(\smallhalf, 1/B)$ in \eqref{rhoxyformula},  i.e. 
to replace $1/B$ by a function $f(B)$.   The resistivity remains quantized on the plateaux,  i.e. equation \eqref{rhoxy} still holds. 
However this modifies the critical values $\Bc_n$ to solutions of $f(\Bc_n) = t_n$.    
   For instance,  changing 
$1/B$ by a constant $\alpha/B$ simply rescales the $\Bc_n$.    More importantly,   replacing $1/B$ with $f(B) = 1/B |\log B|$ makes the small $B$ behavior much  more linear near $B=0$ as previously stated.

\bigskip

\n $\bullet$ ~~ {\bf Deforming the relation between $E_F$ and $B$.}   ~~ The proposed relation $\eqref{EBprop}$ was an over simplification
of the physics,  and was simply taken as a definition of $\Ec_n$.     Clearly the relation $\Ec_n = 1/\Bc_n$ can be deformed, 
which again does not spoil the quantization on the plateaux but modifies $\Ec_n$.    

\begin{figure}[t]
\centering\includegraphics[width=0.9\textwidth]{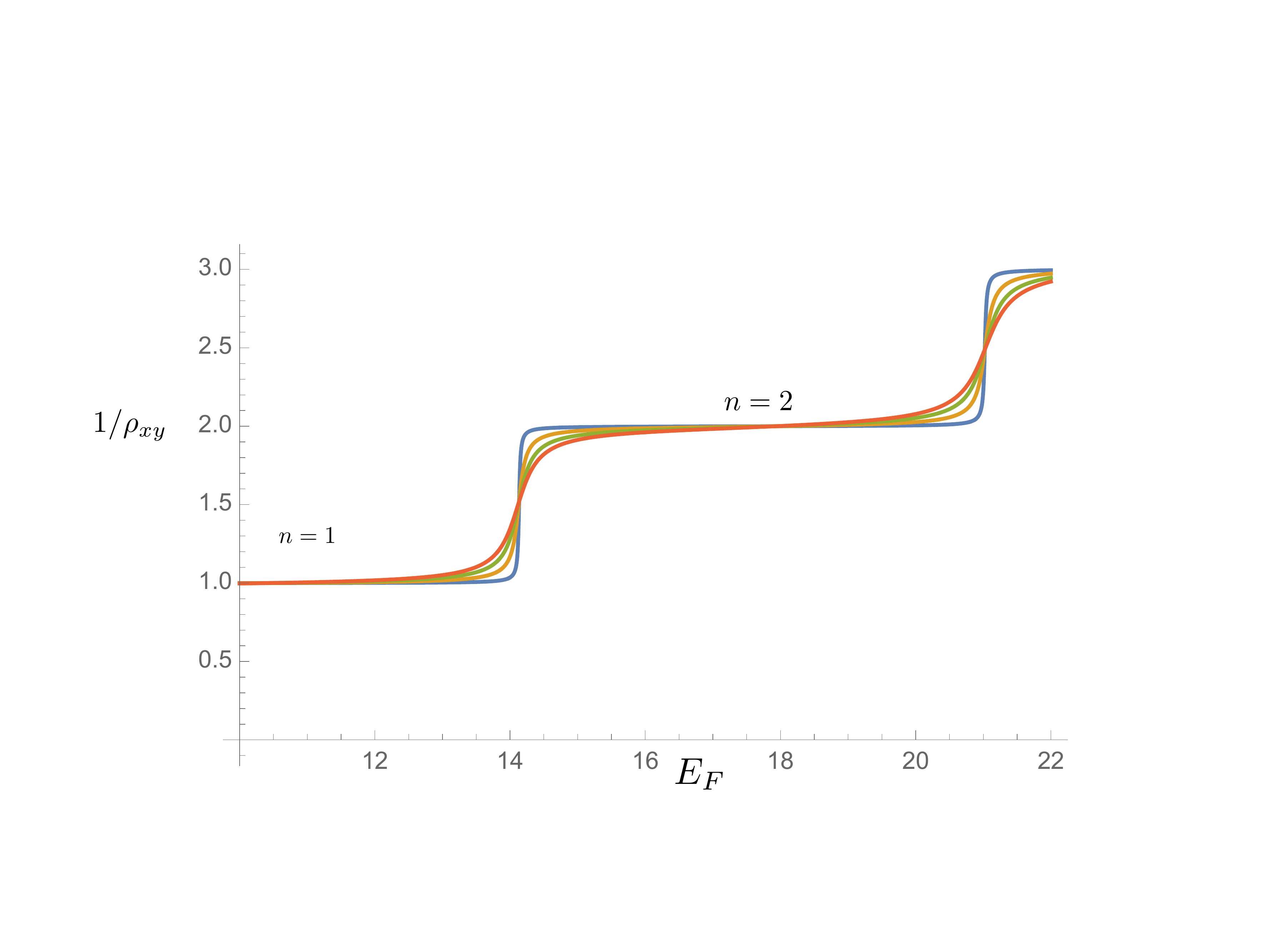}
\caption{The  conductivity $1/\rho_{xy}$ in \eqref{rhoxyPoly} with $\mu = 0.01, 0.05, 0.10, 0.15$.}
 \label{MuPlotKey}
\end{figure}

\bigskip
\n $\bullet$ ~~ {\bf Modeling the effect of finite temperature.} ~~
At zero temperature $\Temp$ the jumps at the transitions in $\rho_{xy}$ are known to be  infinitely sharp,  i.e. are close to step functions. 
At finite temperature these sharp transitions are broadened smoothly and have a finite width.    It is known experimentally that the 
jumps are still centered around the zero temperature $\Bc_n$ but deformed in a smooth way that is symmetric about $\Bc_n$.   
This behavior can be incorporated in a relatively simple way that we now describe.    In finding this deformation we were motivated by 
a deformation of an integral representation of the zeta function that closely resembles adding a chemical potential to the Fermi-Dirac distribution.   Namely:  
\beq
\label{FermiPoly}
- \Gamma (s)\,  {\rm Li}_s (-e^{-\mu}) = \int_0^\infty  d\vep \, \vep^{s-1} \inv{e^{\vep + \mu}  +1 }, ~~~ ~~~~\Re (s) > 0,
\eeq
where ${\rm Li}_s (z)$ is the polylogarithm:
\beq
\label{LiDef}
{\rm Li}_s (z) = \sum_{n=1}^\infty  \frac{ z^n }{n^s} 
\eeq
(analytically continued). 
In the above formula $\vep^{s-2}$ can perhaps be viewed as a kind of density of states.   
The above formula applied to a free gas of fermions would identify $\mu$ as minus the chemical potential divided by the temperature,
however this does not necessarily correspond to the physical situation here.   
It does however  suggest  
to  replace $\zeta (s)$ in \eqref{thetaDef}
by the polylogarithm
\beq
\label{thetaPoly}
\theta_\mu (\sigma, t) = \Im \log \Gamma (s/2)  - t \, \log \sqrt{\pi }+ \arg \, {\rm Li}_{s} (e^{-\mu} ).
\eeq
($s=\sigma+it$).
Note that 
${\rm Li}_s (1) =\zeta (s)$.  
Let us thus consider the deformation:
\beq
\label{rhoxyPoly}
\rho_{xy}  (B,\mu) = \frac{\pi}{\theta_\mu (\smallhalf, 1/B) + 2 \pi } .
\eeq
The resulting resistivity closely captures the broadening of the transitions found in  the experimental data,   as seen in Figure \ref{MuPlotKey}.

It is then interesting to see what \eqref{rhoxyPoly} implies for the de-localization exponent $\nu$.   
At the transition the correlation length diverges $\xi_c \sim |x-x_c|^{-\nu} $ where $x$ could be the magnetic field for instance. 
The broadening of the transition at finite temperature depends on $\nu$.     At zero temperature $\Temp$ the transition is sharp,  thus 
$d\rho_{xy} /dB$ diverges.   It is known  from the physics that  (see for instance \cite{RIMS}): 
\beq
\label{exponentT}
\frac{\d \rho_{xy}}{\d B} \Big\vert_{B=\Bc_n} \sim \Temp^{-\kappa}
\eeq
where 
$\kappa = 1/(\nu \, \zexp)$.   Here $\zexp$ is the dynamical exponent that relates temperature and phase coherence length
$\ell_\phi \sim T^{-1/\zexp}$.    For relativistic systems $\zexp = 1$ and it is known that for the IQHE $\zexp \approx 1$.   
From the formula \eqref{rhoxyPoly} one sees that the sharpness of the transition is controlled by $\mu$,  where the derivative 
in \eqref{exponentT} is infinite when $\mu =0$.     At the $n$-th transition around $\mu =0$,  from the formula \eqref{rhoxyPoly}  the simple  exponent $\kappa =1$ is obtained: 
\beq
\label{exponent}
\frac{\d \rho_{xy}}{\d B} \Big\vert_{B=\Bc_n} = C_n \, \mu^{-\kappa},  ~~~~~\kappa =1 .
\eeq
The exponent $\kappa$ is the same for all $n$ but $C_n$ varies,  for instance 
$C_1 = 17.46,  C_2 =11.59$.
Experimentally $\nu \approx 2.38$ whereas relatively recent analysis of the  Chalker-Coddington network model gives $\nu = 2.59$
\cite{RIMS}.  
Based on the formula \eqref{rhoxyPoly} we can obtain  non-trivial $\kappa \neq 1$ by simply replacing 
$\mu$ by $\mu^\kappa$.

\bigskip
\n $\bullet$ ~~ {\bf A deformation of the actual Riemann zeros.} ~~
There is another  interesting deformation which is rather different than the above which involves deforming the $\smallhalf$ in $\theta(\smallhalf, 1/B)$ in \eqref{rhoxyformula}, as we now explain.   
From $\bar{\chi (s)} = \chi(\bar{s})$ one has 
$\theta (\sigma, -t) = - \theta (\sigma, t)$~ ${\rm mod} ~ 2 \pi$.  
Combined with the functional equation  \eqref{thetaFunc},  one has
\beq
\label{thetasumzero}
\theta(\sigma, t) + \theta(1-\sigma, t)  = 0, ~~~~~{\rm mod} ~ 2 \pi . 
\eeq
 Let us then define 
\beq
\label{thetasum}
\thetasum (\sigma, t) \equiv \smallhalf \(  \theta (\sigma, t) + \theta (1-\sigma, t) \).  
\eeq
  We can thus define what is necessarily an integer $\ncal (\sigma, t)$ for all $\sigma, t$:
\beq
\label{ncal}
\ncal (\sigma, t) = \thetasum(\sigma, t)/\pi +1.
\eeq
Let us then deform $\rho_{xy}$ as follows:
\beq
\label{rhoxyDeformed}
\rho_{xy} (B, \sigma) = \frac{\pi}{\thetasum (\sigma, 1/B) + 2 \pi}.
\eeq

\begin{figure}[t]
\centering\includegraphics[width=.5\textwidth]{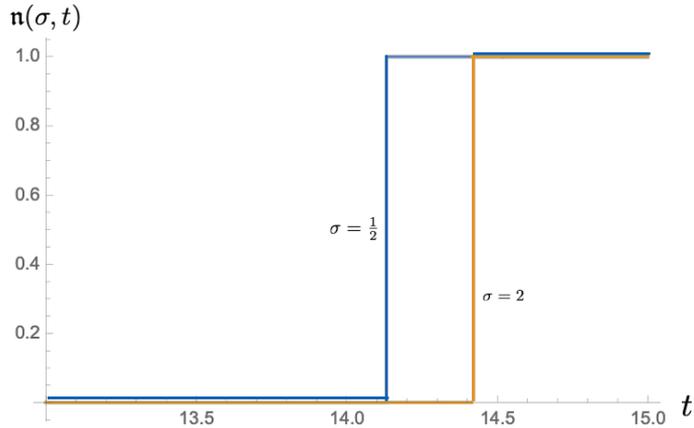}
\caption{$N(t) = \ncal (\sigma = \smallhalf, t)$ (blue line) verses $\ncal (\sigma = 2, t)$ (yellow line) as a function of $t$ around the first zero $t_1=14.1347..$.    For $\ncal (2,t)$ one sees that the plateaux values have not changed, i.e. are still $0$ or $1$, however the transition is deformed away from the actual Riemann zero $t_1 = 14.13..$ to $14.42..$.}
 \label{sigmaDeformation}
\end{figure}

When $\sigma=\smallhalf$,  ~$\ncal (\smallhalf, T)$ equals the number of zeros along the critical line  $N(T)$ in eq. 
\eqref{NofT} (assuming RH).   
We have observed an interesting mathematical property:  
$\ncal (\sigma, T)$ is also equal to $N(T)$ for continuous $\sigma \neq \smallhalf$  but with  small deformations of  the transition values $t_n
\to \hat{t}_n$ that depend on $\sigma$ and $n$.   The deformed $\hat{t}_n$ are clearly no longer zeros of $\zeta$.   One explanation for this property  is that small deformations from $\sigma = \smallhalf$ 
should not change $N(t)$ as long as one is on a plateaux and not too close to the transition,  however we have no proof of this.     
More precisely $\ncal (\sigma, T) = N(T)$ for $T$ deep inside the plateaux away from the transitions.   We have verified this for $n$ up to $1000$.   
In other words this deformation of $\sigma$ from $\smallhalf$ does not change the topological number $n$ on the plateaux,  however
the critical  values $\Ec_n$ are slightly deformed  from $t_n$ in a non-trivial manner.       This is only true for $\sigma$ not very  large,  otherwise the counting is affected by the poles
of $\chi (s)$ at the poles of $\Gamma (s/2)$.   Thus in such a deformation one should limit $ -1 < \sigma < 2 $,  which includes values outside the critical strip.     
We show this numerically in Figure \ref{sigmaDeformation} around the first zero $t_1 = 14.1347..$ for the extreme deformation $\sigma =2$. 
One sees that $t_1$ is deformed to approximately $\hat{t}_1 = 14.42$.  
It is somewhat remarkable that the function $\theta_S (\sigma, t)$ knows about $N(T)$ on the plateaux even though it can be computed from $\zeta$ for values completely {\it outside} the critical strip $\sigma >1$.   
This perhaps has some interesting implications in analytic number theory.   

\bigskip
\n $\bullet$ ~~ {\bf Deformations based on other $L$-functions.} ~ 
On the more exotic side,   in analytic number theory 
there are an infinite number of known $L$-functions that are expected to satisfy the so-called Grand Riemann Hypothesis,  in particular those based on Dirichlet characters or on modular forms \cite{Apostol}.    Replacing $\theta$ in \eqref{rhoxyformula} with the argument of the completion of such an $L$-function analogous to \eqref{chidef},  which is known to also involve the gamma function \cite{Apostol}, 
  one still has a robust quantization for $\rho_{xy}$,  however the $t_n$  and thus the $\Bc_n$ and $\Ec_n$ are different in a non-smooth way.

\bigskip

\section{Closing remarks}

We have proposed a phenomenological formula for the transverse resistivity for the IQHE built from the gamma and zeta functions 
which appear to capture the main physical properties at least qualitatively.     The physics is very speculative:   we emphasize once again that we have not performed any computation of the resistivity in any specific quantum many-body problem whatsoever,   thus the main open question is whether a physical model exists that has a resistivity 
corresponding to our proposed formula.    If we simply assume the formula we proposed for $\rho_{xy}$,   then the non-trivial Riemann zeros play an essential role,  and this perhaps offers a new perspective on the Riemann Hypothesis which we have partially explored. 
For instance,  if the RH were false,  then not all integers $n$ in the quantization \eqref{rhoxy} would be physically  realized.    
All of the pure mathematics we have used is well-known,  except for the discussion surrounding \eqref{rhoxyDeformed}.   

A  common link between the IQHE and the Riemann zeros perhaps comes from  random matrix theory,  since the latter has been applied to both the Riemann zeros and to disordered systems.   This is an appealing connection,   to be contrasted to the hamiltonians proposed toward a realization of 
the Hilbert-P\'olya idea.     Proposals such as  $H= xp$ and variations  are not  random hamiltonians;    instead the randomness  of the zeros is attributed to the chaotic behavior of such hamiltonians \cite{BerryK},  which is very different than a particle moving in a random landscape,  such as in the IQHE.       
In fact,  relatively recently,  random matrix theory was applied to a study of the extreme values of the zeta function on the critical line
by making an analogy with the so-called freezing transition in disordered landscapes \cite{Fyodorov1,Fyodorov2},  
and such a  freezing transition is expected to play a role in the IQHE in order to understand  its multi-fractal properties.

\section{Acknowledgements}
We wish to thank Giuseppe Mussardo and Germ\`an Sierra for discussions.

\appendix

\section{Some basic properties of the Riemann zeta function}

In this Appendix we summarize some of the  fundamental properties of the zeta function that we need  \cite{Edwards}.
Adopting standard notations in analytic number theory,  throughout $s=\sigma + i t$ is a complex variable.  

The zeta function was originally defined by the series
\beq
\label{zetadef}
\zeta (s) = \sum_{n=1}^\infty  \inv{n^s} , ~~~~~~\sigma > 1
\eeq
which converges for $\sigma>1$.  
It can be analytically continued to the entire complex plane where it has a simple pole at $s=1$.
It has trivial zeros at $s = -2m$, $m=0,1, 2, \ldots$.    It is known to have an  infinite number of zeros inside the ``critical strip" 
$0< \sigma < 1$.   It is also known there are an infinite number of zeros along the ``critical line" $\sigma = \smallhalf$.  
The Riemann Hypothesis is the statement that the latter are the only zeros inside the critical strip.  
We label those on the upper critical line as  $\rho_n = \smallhalf + i \,t_n$, $t_n > 0$: 
\beq
\label{zeros}
\zeta(\rho_n) =0,  ~~~ \rho_n = \smallhalf + i\, t_n ,   ~~~~n= 1,2,3, \ldots
\eeq
The property $\bar{\zeta (s)} = \zeta (\bar{s})$ implies $\bar{\rho_n} = \smallhalf -i t_n $ is also a zero. 

It is perhaps worth pointing out that Riemann first performed the analytic continuation based on 
an integral representation for $\sigma > 1$ involving the Bose-Einstein distribution:
\beq
\label{IntRepBose}
\Gamma (s) \, \zeta (s) = \int_0^\infty \, d\vep \,  \vep^{s-2} \, \(  \frac{\vep}{e^{\vep}-1}\), ~~~~~\Re (s) > 1.
\eeq
The integration contour can be deformed such that $\zeta (s)$ is defined everywhere in the complex plane, except at the pole at $s=1$.
There exists another integral representation involving the Fermi-Dirac distribution: 
\beq
\label{IntRepFermi}
\(1-2^{1-s} \) \, \Gamma (s) \, \zeta (s) = \int_0^\infty \, d\vep \,  \vep^{s-2} \, \(  \frac{\vep}{ e^{\vep} +1}\), ~~~~~\Re (s) > 0,
\eeq
which motivated some results in Section IV.

Let us define a completed zeta function as follows:\footnote{In Riemann's original paper,  he worked with 
$\xi (s) = s(1-s) \chi(s)/2$ in order to remove the pole at $s=1$.   This is not necessary for our purposes.}
\beq
\label{chidef}
\chi (s) \equiv \pi^{-s/2} \,  \Gamma (s/2)  \, \zeta (s) .
\eeq
It satisfies the  important functional equation:
\beq
\label{funceqn}
\chi (s) = \chi (1-s) .
\eeq
This implies that zeros off the critical line necessarily come in pairs symmetric about $\smallhalf$;  namely 
 if $\rhozero = \sigmazero + i \tzero$ is a zero,  then so is 
$\rhozero = 1-\sigmazero + i \tzero$. 

The angle $\theta(\sigma, t)$ defined in \eqref{thetaDef} is simply its argument:
\beq
\label{thetaarg}
\theta (\sigma, t ) = \arg \, \chi (\sigma + it ).
\eeq
One property we  will need is 
\beq
\label{thetaFunc}
\theta (\sigma, t)  =  \theta (1-\sigma , -t) 
\eeq
which follows from the functional equation.   
From this,  one can see that $\arg \, \zeta(s)$ behaves very differently to the right verses to the left of the critical line.
Using the Stirling formula, in the limit of large $t$ one has
\beq
\label{argrelations}
\arg \, \zeta ( 1-\sigma + i t) = - \arg \, \zeta (\sigma + it ) - t \log \( t/2 \pi e \) + \frac{\pi}{4} +\frac{(6 \sigma^2 - 6 \sigma +1 )}{12 t} + \CO(1/t^3) 
\eeq
(${\rm mod} ~ 2 \pi$).

\begin{figure}[t]
\centering\includegraphics[width=.7\textwidth]{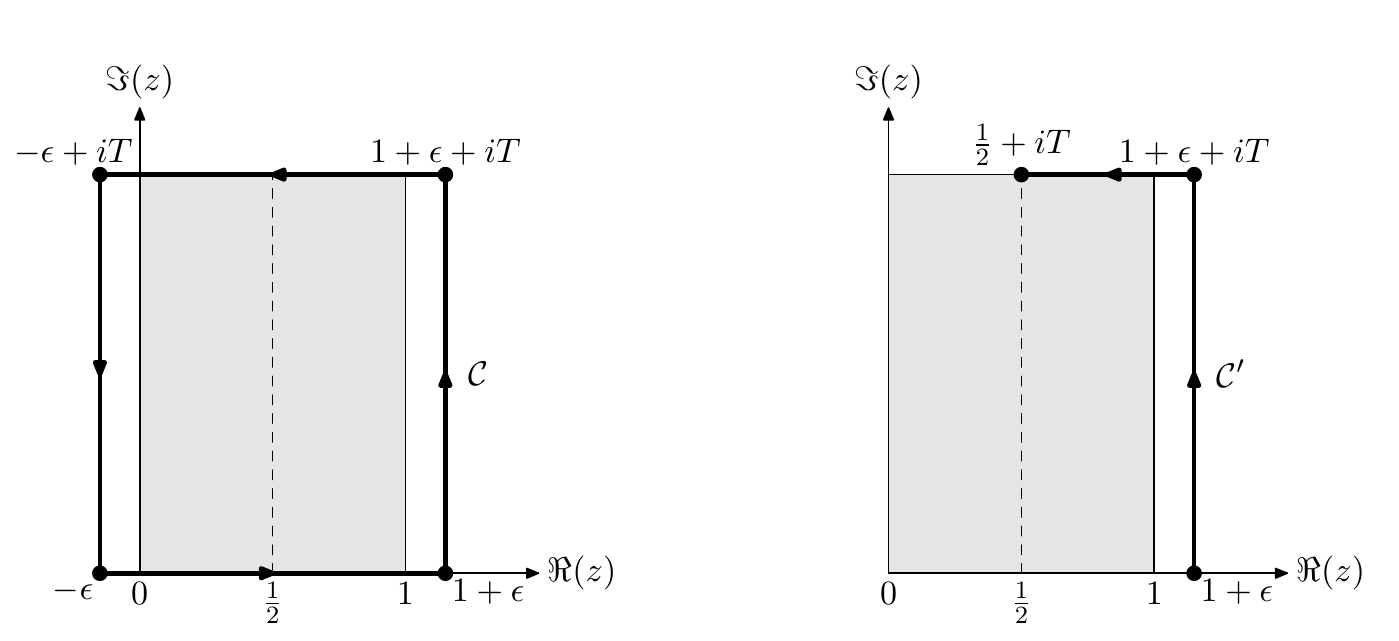}
\caption{{\bf Left:}.  The contour $\CC$ for the argument principle used to determine the number of zeros
in the critical strip with $0<t<T$ referred to as $N(T)$ in \eqref{NofT}. (In this figure, the complex variable $s$ is denoted as $z$.)
The critical line $\sigma = \smallhalf$  splits the rectangle in half.  
{\bf Right: }  It is sufficient to double the result for the contour $\CC'$ due to \eqref{thetasumzero}.}
 \label{argFigure}
\end{figure}

Cauchy's argument principle determines the number of zeros in the critical strip with ordinate $0<t< T$, commonly  referred to as $N(T)$ in the mathematics literature. 
Recall 
\beq
\label{argPrinciple}
\inv{2 \pi i} \oint_\CC \frac{\chi'(s)}{\chi(s)} \, ds = N_{{\rm zeros}} - N_{{\rm poles}}
\eeq
where the number of zeros $N_{{\rm zeros}}$ includes their multiplicity,  
and the number of zeros $N_{{\rm poles}}$ includes their order.   One has 
\beq
\label{argprintheta}
\inv{2 \pi i} \oint_\CC \frac{\chi'(s)}{\chi(s)} \, ds 
=  \inv{2 \pi} \oint_\CC \,  d \theta (s) .
\eeq  
The contour $\CC$ chosen is shown in Figure \ref{argFigure}.
One need only consider the part of the contour $\CC'$ to the right of the critical line by virtue of \eqref{thetaFunc}.
In this way one obtains the formula \eqref{NofT} for $N(T)$.   
The shift by $1$ is due to the simple pole at $s=1$.  
This formula is only valid if $T$ is not the ordinate of a zero.

\end{document}